\documentclass[12pt]{article}
\usepackage{epsfig}

\def\beq{\begin{equation}}
\def\eeq#1{\label{#1}\end{equation}}
\def\eeqn{\end{equation}}

\def\beqa{\begin{eqnarray}}
\def\eeqa#1{\label{#1}\end{eqnarray}}
\def\eeqan{\end{eqnarray}}

\let\bar=\overbar

\def\Dslash{\not{\hbox{\kern-4pt $D$}}}
\def\dslash{\not{\hbox{\kern-2pt $\del$}}}

\def\msb{{\bar{\ssstyle M \kern -1pt S}}}

\def\Title#1{\begin{center} {\Large {\bf #1} } \end{center}}

\begin{document}

\Title{Single top-quark production with the ATLAS detector
in pp collisions at $\sqrt{s}$~=~7~TeV}

\bigskip\bigskip


\begin{raggedright}  

{\it Reinhard Schwienhorst\footnote{For the ATLAS collaboration}\index{Schwienhorst, R.}\\
Department of Physics and Astronomy\\
Michigan State University\\
East Lansing, MI 48824, USA}
\bigskip\bigskip
\end{raggedright}

\abstract{
The ATLAS experiment at the LHC at CERN has analyzed 2010 and 2011 data
looking for electroweak production of single top quarks in the lepton+jets
and di-lepton final states. The production
cross section for the $t$-channel process is measured to be 
$76^{+41}_{-21}$~pb using 156~pb$^{-1}$ of 2011 data.
A first limit is set on the $Wt$~associated production process using
lepton+jets and di-lepton events. The 95\% CL upper limit on the $Wt$ 
production cross section is 158~pb using 35~pb$^{-1}$ of 2010 data.
}

\section{Introduction}

\begin{figure}[htb]
\begin{center}
\epsfig{file=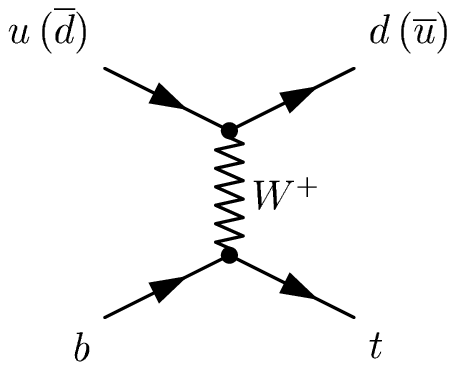,height=1.3in} \hspace{0.3in}
\epsfig{file=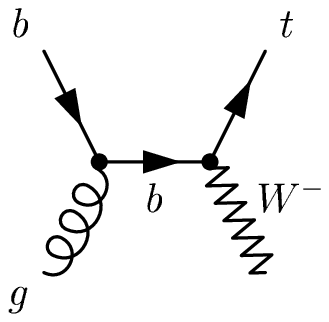,height=1.3in} \hspace{0.3in}
\epsfig{file=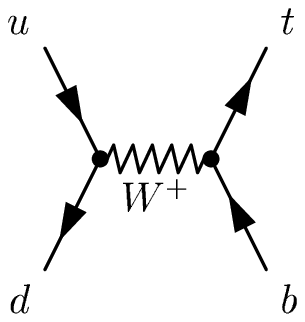,height=1.3in}
\caption{Feynman diagrams for single top quark production in the $t$-channel,
$Wt$ associated production and $s$-channel mode.
}
\label{fig:feynman}
\end{center}
\end{figure}

Single top quark production in the SM is the electroweak production of a 
single top quark at a hadron collider. A measurement of the single top quark 
production cross section provides a direct measurement of the the quark mixing 
matrix element $|V_{tb}|$~\cite{singletop-vtb-jikia}. It also serves as a probe 
of the $Wtb$ coupling~\cite{Chen:2005vr,Abazov:2009ky}
and is sensitive to several models of new physics~\cite{Tait:2000sh}.
Single top quark production at the LHC proceeds via three separate
modes shown in Fig.~\ref{fig:feynman}, each resulting in a unique final 
state: the $t$-channel exchange of a virtual $W$~boson between a light quark 
line and a heavy quark 
line, the $Wt$~associated production of a top quark and a $W$~boson, and the
$s$-channel production and decay of a virtual $W$~boson.
The predicted cross sections for single top production at NLO are
66.2~pb for the $t$-channel, 14.6~pb for $Wt$ associated production and
4.3~pb for the $s$-channel~\cite{Torrielli:2010aw}.

~

Single top quark production was observed in 2009 at the 
Tevatron~\cite{Aaltonen:2009jj,Abazov:2009ii,Group:2009qk}, and the $t$-channel 
mode was
also observed by the D0 collaboration~\cite{Abazov:2011rz}.

This note presents a measurement of the $t$-channel cross section using 
156~pb$^{-1}$ of 2011 7~TeV ATLAS lepton+jets data~\cite{ATLAS-CONF-2011-088} 
and a search for $Wt$~associated 
production using 35~pb$^{-1}$ of 2010 7~TeV ATLAS data in the lepton+jets and 
di-lepton channels~\cite{ATLAS-CONF-2011-027}.

\section{T-channel analysis}
~
Single top-quark events are selected in the final state containing one 
isolated lepton
(electron or muon with $p_T>25$~GeV), missing transverse energy 
($E_T^{miss}>25$~GeV) and two jets ($p_T>25$~GeV), exactly one of which is 
required to be $b$-tagged. 

The main backgrounds to this signature are from $W$+jets 
production, QCD multijet events, and top quark pairs. 
Smaller backgrounds are due to $Z$+jets and diboson production. 
~

\begin{figure}[htb]
\begin{center}
\epsfig{file=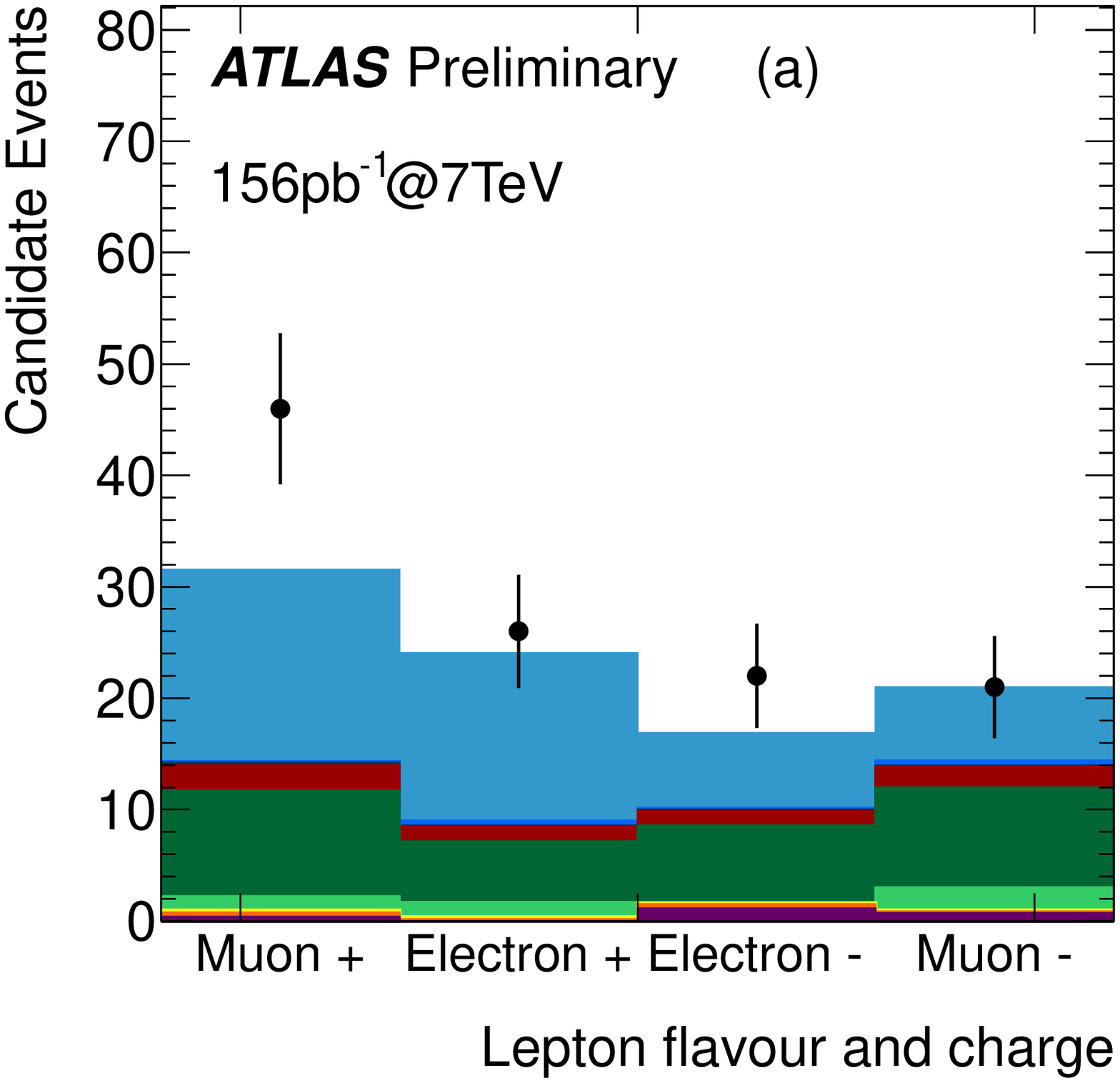,height=2.0in}
\hspace{-0.2in}
\epsfig{file=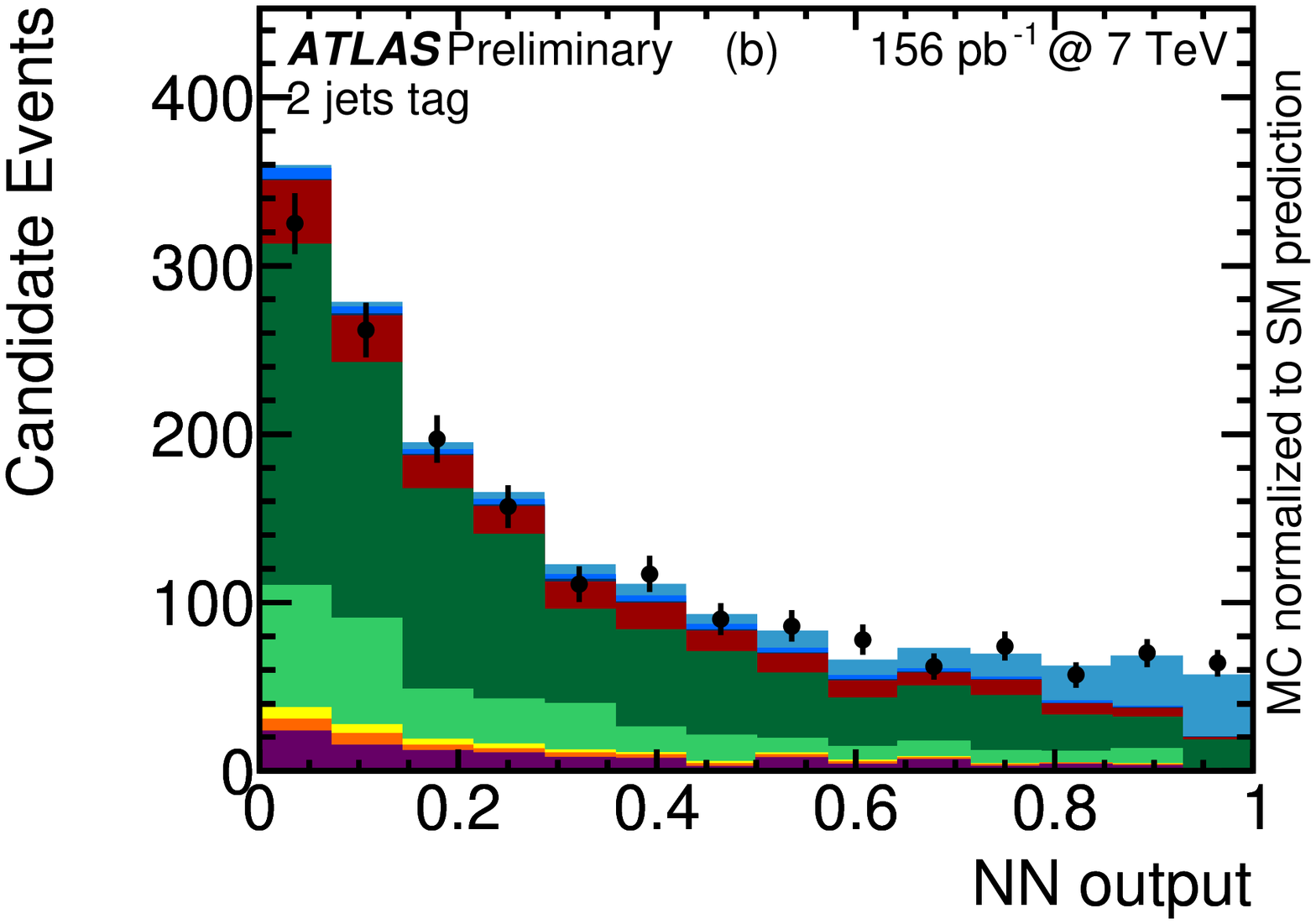,height=1.7in}
\hspace{-0.1in}
\epsfig{file=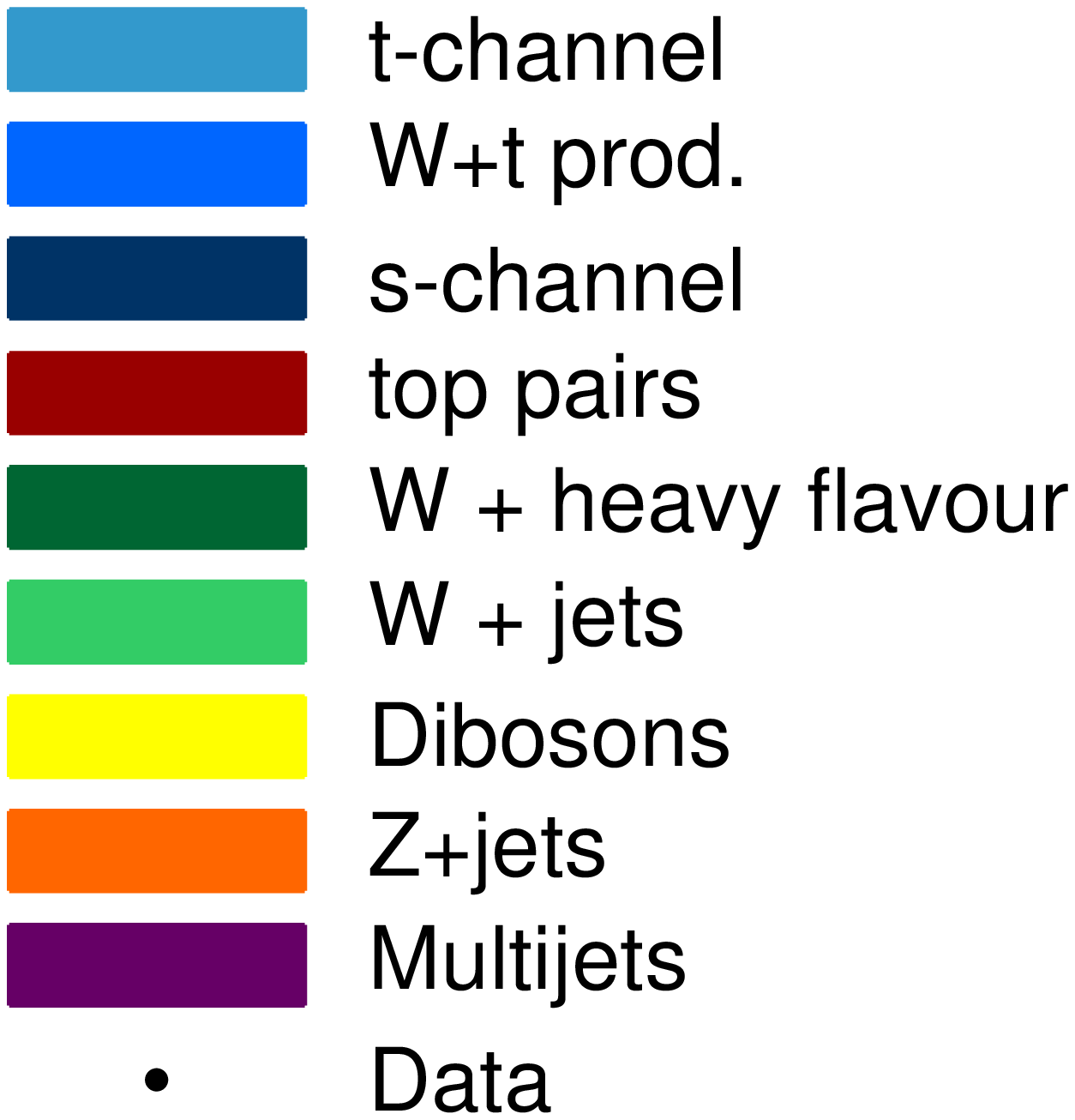,height=1.1in}
\caption{(a) Single top $t$-channel cut-based events separated by lepton flavour 
and charge and (b) neural network output distribution.
}
\label{fig:cutnn}
\end{center}
\end{figure}

The $t$-channel analysis is done both using a cut-based approach and 
a multivariate approach employing a Neural Network (NN). The cut-based 
approach requires the light quark jet (non $b$-tagged jet) to be forward in 
pseudo-rapidity ($\eta>2.0$), 
the reconstructed top quark mass to be between 140~GeV and 190~GeV,
the sum of the transverse energies of all objects to be larger than 180~GeV,
the pseudo-rapidity difference between the two jets to be large ($\Delta\eta>2.0$),
and the $b$-tagged jet to be central ($\eta_b<2.0$).
The distribution of the selected events as a function of lepton flavour and
charge is shown in Fig.~\ref{fig:cutnn}(a). 
~

The NN analysis of $t$-channel events uses 22 
discriminating variables. The most important variable is the reconstructed 
top-quark mass, followed by the invariant mass of the two jets and the 
pseudo-rapidity of the light quark jet. The resulting NN output
distribution is shown in Fig.~\ref{fig:cutnn}(b). For the final statistical
analysis a cut is made at a NN value of 0.86. 

The cross section measurement is extracted using a frequentist approach with
profiling of systematic uncertainties. Systematic uncertainties are larger than
statistical uncertainties, with the largest contributions to the systematic
uncertainty due to the signal modeling, jet energy calibration, 
$b$-tag modeling and background normalization.
The $t$-channel cross section is measured to be $97^{+54}_{-30}$~pb by the
cut-based analysis and $76^{+41}_{-21}$~pb by the neural network analysis. Both
are consistent with the SM expectation. The observed significances are 6.1 
standard deviations for the cut-based analysis and 6.2 standard deviations 
for the neural network analysis.

\section{Wt analysis}
~
Single top-quark $Wt$ events are selected in two final states, one containing
two isolated leptons and one jet and the other containing one lepton and two to four
jets. The lepton+jets analysis has similar backgrounds as the $t$-channel analysis
and uses the same object selection and background estimation methods. It requires
the $b$-tagged jet to have $p_T>35$~GeV and the angular separation between the two
leading jets to be less than 2.5.

\begin{figure}[htb]
\begin{center}
\epsfig{file=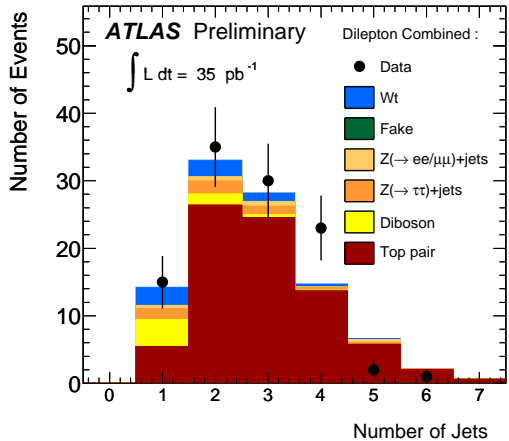,height=2.5in}
\vspace{-0.2in}
\caption{Jet-multiplicity for the single top $Wt$ dilepton analysis.
}
\label{fig:njets}
\end{center}
\end{figure}

The dilepton analysis requires two opposite-charge leptons with $p_T>20$~GeV,
one jet with $E_T>20$~GeV and $E_T^{miss}>50$~GeV. The main backgrounds to this
event signature are from top quark pair di-lepton events and 
Drell-Yan production of $Z/\gamma$+jets. Smaller contributions come from
$Z\rightarrow\tau\tau$ events, dibosons, $W$+jets and multijet events. The 
jet-multiplicity distribution for dilepton events is shown in Fig.~\ref{fig:njets}.
~

The observed data are consistent with the background-only expectation in both
the dilepton and the lepton+jets channels for the $Wt$ analysis. An upper limit
is set on the $Wt$ production cross section using a frequentist approach with
profiling of systematic uncertainties. The dominant sources of systematic
uncertainty are the top quark pair production modeling and the background
normalization. The combined
limit on the $Wt$ production cross section at the 95\% confidence level (CL)
is 158~pb.

\section{Summary}
~
The ATLAS experiment has observed single top quark production in the $t$-channel
at the LHC. This is the first observation of single top at the LHC and 
complements Tevatron results. ATLAS has also set a first upper limit on 
$Wt$ associated production.



\end{document}